\pdfoutput=1
\documentclass[preprint,
 amsmath,amssymb,
 aps,
]{revtex4-2}

\usepackage{graphicx}
\usepackage{dcolumn}
\usepackage{bm}
\usepackage{tabularx}
\usepackage{multirow}
\usepackage{booktabs}
\usepackage{siunitx}
\usepackage{subfig}
\usepackage{hyperref}
\usepackage{multirow}

\begin{document}


\title{Prospects for measuring electroweak production of $Z\gamma\gamma$ and 2 jets at the LHC}

\author{Ran Ding}
\email{ran.ding@cern.ch}
\affiliation{School of Physics and State Key Laboratory of Nuclear Physics and Technology, Peking University, Beijing, 100871, China}

\author{Ruobing Jiang}
\affiliation{School of Physics and State Key Laboratory of Nuclear Physics and Technology, Peking University, Beijing, 100871, China}

\author{Tianyi Yang}
\affiliation{School of Physics and State Key Laboratory of Nuclear Physics and Technology, Peking University, Beijing, 100871, China}

\author{Andrew Levin}
\affiliation{School of Physics and State Key Laboratory of Nuclear Physics and Technology, Peking University, Beijing, 100871, China}

\author{Qiang Li}
\affiliation{School of Physics and State Key Laboratory of Nuclear Physics and Technology, Peking University, Beijing, 100871, China}

\date{\today}

\begin{abstract}
Vector boson scattering (VBS) serves as a powerful channel for probing the Standard Model, particularly the electroweak symmetry breaking mechanism. Currently, studies of VBS mainly focus on $2 \to 2$ scattering. In this study, we investigate the $2\to 3$ VBS process of $p p \to Z\gamma\gamma + 2~\text{jets}$ through Monte Carlo simulations, including signal generation and background analysis. The signal significance is evaluated across different phase-space regions. With an integrated luminosity of 500 fb$^{-1}$, the signal significance can reach about $4.5~\sigma$. These results suggest that, given the ongoing release of the LHC Run 3 dataset, there will be promising opportunities to explore and potentially discover a series of $2\to 3$ VBS processes, starting with the $Z\gamma\gamma$ channel.
\end{abstract}

\maketitle

\newpage

\section{Introduction}
\label{sec:introduction}

Following the observation of the Higgs boson at the Large Hadron Collider (LHC) in 2012~\cite{ATLAS:2012yve,CMS:2012qbp}, the ATLAS~\cite{ATLAS:2008xda} and CMS~\cite{CMS:2008xjf} collaborations have conducted both precision measurements of Higgs boson properties and detailed studies of electroweak (EWK) interactions~\cite{CMS:2022dwd,ATLAS:2022vkf}. Crucially, vector boson scattering (VBS) stands out among these as a key channel for probing EWK interactions~\cite{BuarqueFranzosi:2021wrv}. These processes involve multiple self-coupling vertices among the EWK bosons $W$, $Z$, and $\gamma$, thus offering direct sensitivity to the electroweak symmetry breaking mechanism~\cite{Englert:1964et}.

Currently, studies of VBS processes mainly focus on $2 \to 2$ scattering, with outgoing vector bosons types including $WW$~\cite{CMS:2020gfh,ATLAS:2016snd}, $ZZ$~\cite{CMS:2020fqz,ATLAS:2020nlt}, $WZ$~\cite{CMS:2020gfh,ATLAS:2018mxa}, $W\gamma$~\cite{CMS:2020ypo}, and $Z\gamma$~\cite{CMS:2020ioi,ATLAS:2019qhm}. These analyses are largely based on data from LHC Run 2, with a total integrated luminosity of about 140~fb$^{-1}$ and a center-of-mass energy of $\sqrt{s} = 13\ \text{TeV}$. In recent years, the LHC has been conducting Run 3 data-taking at $\sqrt{s} = 13.6~\text{TeV}$, and the combined luminosity of both Run 2 and Run 3 is expected to exceed 400~fb$^{-1}$~\cite{lhc_lumi}. With this large dataset, it will be possible not only to improve the precision of previous measurements but also to strongly support the search, discovery, and cross-section measurement of $2 \to 3$ VBS processes.

Though no officially released results for $2 \to 3$ VBS processes have been reported, studies on the production of triple vector bosons via vector boson fusion (VBF) are relatively common. Examples include final states with only weak bosons, such as $WWW$~\cite{CMS:2020hjs}, $WWZ$~\cite{ATLAS:2024nab,CMS:2025hlu}, $WZZ$~\cite{ATLAS:2024nab}, and $ZZZ$~\cite{ATLAS:2024nab}, as well as processes involving photons like $WW\gamma$~\cite{ATLAS:2017bon,CMS:2023rcv}, $WZ\gamma$~\cite{ATLAS:2017bon,CMS:2025oey}, $W\gamma\gamma$~\cite{ATLAS:2023avk,CMS:2021jji}, and $Z\gamma\gamma$~\cite{CMS:2021jji,ATLAS:2022wmu}. Although the phase space regions examined in these analyses differ considerably from those typical of VBS processes, their well-developed methodologies offer valuable insights for future VBS studies. In summary, newly available data, together with established analysis techniques from diboson VBS and triboson VBF studies, provide a solid foundation for the next stage of $2 \to 3$ VBS research.

\begin{figure}[!h]
    \centering
    \includegraphics[width=0.45\linewidth]{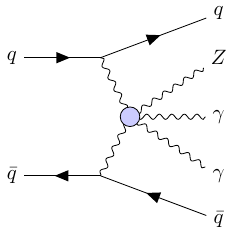}
    \caption{Feynman diagram of VBS $Z\gamma\gamma$ production.}
    \label{fig:Feynman_ZGGVBS}
\end{figure}

In this study, we investigate the VBS $Z\gamma\gamma$ process, with the corresponding example Feynman diagram as shown in Figure \ref{fig:Feynman_ZGGVBS}. It naturally mixes and interferes with other QCD-induced and EWK $\text{pp} \to Z\gamma\gamma + 2~\text{jets}$ diagrams, which is a common challenge in VBS analyses. Following previous studies, we define the signal as the full EWK component of $\text{pp} \to Z\gamma\gamma + 2~\text{jets}$, while focus primarily on the VBS-enriched phase space region~\cite{BuarqueFranzosi:2021wrv}. The QCD-induced processes are treated as a major background, and the interference between EWK and QCD contributions is examined in Section \ref{sec:inter}. For simplicity, only events with $Z \to \ell^+\ell^-$ ($\ell = e, \mu$) are considered.

This paper is structured as follows. After this introduction, Section~\ref{sec:simulation} details the signal definition, classifies relevant backgrounds, and describes the Monte Carlo (MC) simulation setup. Section~\ref{sec:signalselection} presents the event selection, then gives the comparison of the distributions among signal and background. Section~\ref{sec:yieldandsignificance} shows the final signal and background event yields in different regions, together with the estimated signal significance. A summary of the study is provided in Section~\ref{sec:summary}.

\section{Process classification and simulation}
\label{sec:simulation}

\subsection{Process classification}

As introduced in Section~\ref{sec:introduction}, the signal under study is the EWK-induced process $\text{pp} \to Z\gamma\gamma + 2\ \text{jets}$, with $Z \to \ell^+\ell^-$ ($\ell = e, \mu$). 
In the following study, we further categorize the signal process into two scenarios: the first is the EWK $Z\gamma\gamma + 2~\text{jets}$ process, where an on-shell Z boson subsequently decays to a lepton pair; the second is referred to as the inclusive process, namely the EWK production of $\ell^+\ell^-\gamma\gamma + 2~\text{jets}$.
Typical VBS kinematics are characterized by two jets with large rapidity gap, $\left|\Delta\eta_{jj}\right|$, and large invariant mass, $m_{jj}$, which can be used to suppress both backgrounds and non-VBS electroweak contributions.

In typical real-world experiments, the backgrounds for VBS processes associated with $W$, $Z$, and $\gamma$ primarily include the following categories~\cite{CMS:2020ypo,CMS:2020ioi}:
\begin{itemize}
    \item QCD-induced processes,
    \item Fake photon backgrounds,
    \item Fake lepton backgrounds.
\end{itemize}

In our study, the QCD-induced background refers specifically to the $Z\gamma\gamma + n\text{ jets}$ process produced via QCD interactions. Additionally, the fake photon or lepton backgrounds, referring to cases where these objects do not originate from hard scattering but result from soft processes or misidentification, must also be carefully evaluated. In processes involving $\gamma$ and $W$ bosons, both fake photons and fake leptons typically contribute significantly. In contrast, for processes with only $\gamma$ and $Z$ bosons, the dominant background comes from fake photons. These primarily arise from jet–photon misidentification, which occurs when a high-momentum $\pi^0$ decays into a photon pair and the resulting jet is reconstructed as a photon. In experimental analysis, such backgrounds are commonly estimated using data-driven methods~\cite{CMS:2021jji}. However, in our simulation-based study, the absence of real data prevents the use of this approach. Instead, we model the fake photon contribution using simulated $Z\gamma + n\ \text{jets}$ events. The $Z + n\ \text{jets}$ process is preliminarily neglected, as the probability of double misidentification is substantially lower than that of single misidentification. Other minor backgrounds,  including $t\bar{t}\gamma$ and $tW\gamma$, are also estimated through direct simulation. Though these processes yield at least one fake photon, they are not covered by the $Z\gamma$ simulation.

\begin{table}
\caption{Summary of the signal and background processes investigated in this study.}
\label{tab:process_summary}
\begin{ruledtabular}
\begin{tabularx}{\textwidth}{l l}
\textbf{Category}  & \textbf{Simulated Syntax}\\
\midrule
Signal& \texttt{p p > z a a j j QCD=0, z > l+ l-}\\
\cmidrule{1-2}
Signal (inclusive)& \texttt{p p > l+ l- a a j j QCD=0}\\
\midrule
\multirow{3}{*}{QCD $Z\gamma\gamma$}&\texttt{p p > l+ l- a a}\\
 & \texttt{p p > l+ l- a a j}\\
&\texttt{p p > l+ l- a a j j}\\
\cmidrule{1-2}
\multirow{3}{*}{QCD $Z\gamma$} & \texttt{p p > l+ l- a}\\
& \texttt{p p > l+ l- a j}\\
& \texttt{p p > l+ l- a j j}\\
\cmidrule{1-2}
EWK $Z\gamma$ & \texttt{p p > l+ l- a j j QCD=0} \\
\cmidrule{1-2}
$t\bar{t}\gamma$ & \texttt{p p > t t\char`~ ~a,  (t > w+ b , w+ > l+ vl) , (t\char`~ > w- b\char`~ , w- > l- vl\char`~  )}\\
\cmidrule{1-2}
\multirow{2}{*}{$tW\gamma$} & \texttt{p p > t\char`~ ~w+ a , w+ > l+ vl , ( t\char`~ > w- b\char`~, w- > l- vl\char`~ ~)}\\
& \texttt{p p > t w- a , w- > l- vl\char`~, ( t > w+ b , w+ > l+ vl )}\\
\end{tabularx}
\end{ruledtabular}
\end{table}

\subsection{Events simulation}

The production cross-sections and MC event samples for both signal and background processes are generated using MadGraph5\_aMC@NLO (MG5) v3.6.3~\cite{Madgraph} at Leading Order (LO). The MG5 simulation is performed at $\sqrt{s} = 13.6~\text{TeV}$ with generator-level cuts, requiring $p_T^{\ell} > 10~\mathrm{GeV}$, $p_T^{\gamma} > 15~\mathrm{GeV}$, $p_T^{j} > 20~\mathrm{GeV}$, $|\eta^{\ell,\gamma}| < 2.5$, $|\eta^{j}| < 5$, $\Delta R > 0.4$, and $m_{\ell^+\ell^-}>50~\mathrm{GeV}$. Then parton showering and hadronization are carried out using \textsc{Pythia8} (PY8)~\cite{Pythia} with its default models. The detector response is then simulated using \textsc{Delphes}~\cite{Delphes} v3.5, employing \href{https://github.com/delphes/delphes/blob/master/cards/delphes_card_CMS.tcl}{the built-in CMS detector card}. Specially for QCD $Z\gamma$ and QCD $Z\gamma\gamma$, we consider events with up to 2 jets and merge them using the MLM scheme.  Table \ref{tab:process_summary} gives a summary of the processes and their simulated syntax in MG5 investigated in this study.

\subsection{Interference study}
\label{sec:inter}
The interference term between the Signal and QCD $Z\gamma\gamma$ can be evaluated by simulating through MG5 with requiring \texttt{QCD\^{}2==2}. Table \ref{tab:interferencestudy} shows the cross-sections comparison among the signal, QCD and interference. The interference contribution of $Z\gamma\gamma+2~\text{jets}$ is more than an order of magnitude lower than the EWK signal, so it can be safely neglected in this study.

\begin{table}[h]
\caption{Comparison of the cross-sections among the signal, QCD and interference (fb).}
\label{tab:interferencestudy}
\begin{ruledtabular}
\begin{tabular}{c c c c }
Signal &Signal (inclusive)& QCD $Z\gamma\gamma$ & Interference\\
\midrule
2.63E-1&4.26E-1&2.30E+1&2.44E-2\\
\end{tabular}
\end{ruledtabular}
\end{table}

\section{Signal selection and distribution}
\label{sec:signalselection}
For events that have undergone Delphes simulation, the signal selection and background suppression procedures are applied sequentially. A signal candidate should contain at least two good photons, one pair of opposite-sign same-flavor good leptons, and two good jets. The definitions of good particles are $p_T^{\ell} > 15~\mathrm{GeV}$, $p_T^{\gamma} > 20~\mathrm{GeV}$, $p_T^{j} > 25~\mathrm{GeV}$, $|\eta^{\ell,\gamma}| < 2.5$, and $|\eta^{j}| < 5$. These criteria can effectively suppress soft photons and leptons produced during hadronization.

It is worth noting that since the photons can also be generated during PY8 simulation, there might be double counting effects between the processes QCD $Z\gamma$ and QCD $Z\gamma\gamma$, similarly EWK $Z\gamma$ and the Signal. By tracking the parent particles of the reconstructed-level photons, they can be categorized into two types: those originating from quarks or bosons, referred to as hard photons, and those originating from leptons or hadrons, known as soft photons. For the $Z\gamma\gamma$ processes, ignoring the loss in detection efficiency, it should contain two hard photons. Therefore, we require those $Z\gamma$ events that they must not contain two or more hard photons, thereby avoiding overlap with the $Z\gamma\gamma$ events.

\begin{figure*}[!h]
    \centering
    \subfloat[$M_{\ell^+\ell^-}$]
    {\includegraphics[width=.495\textwidth]{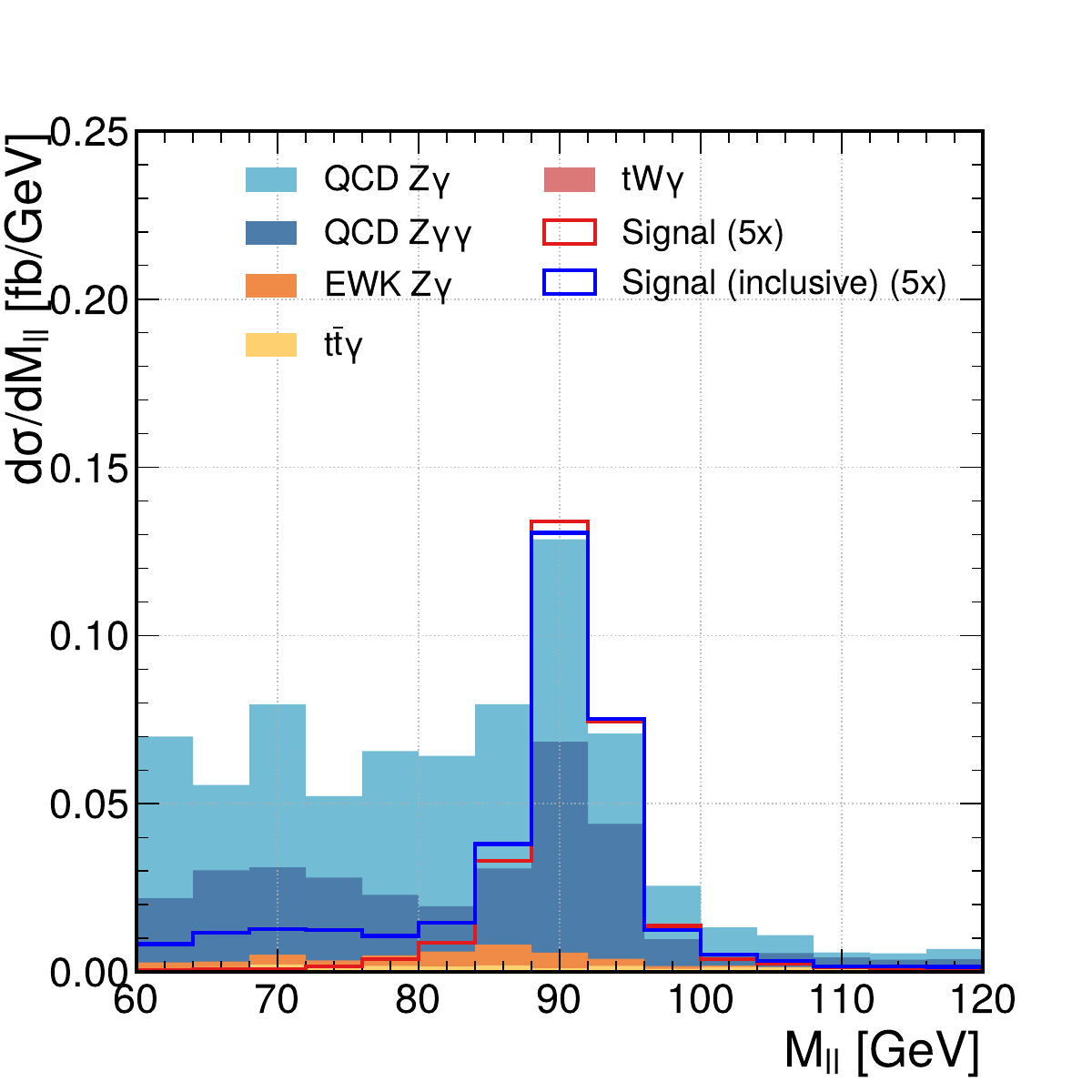}\label{fig:id1_1}}
    \subfloat[$p_{\text{T}}^{\gamma,\text{trailing}}$]
    {\includegraphics[width=.495\textwidth]{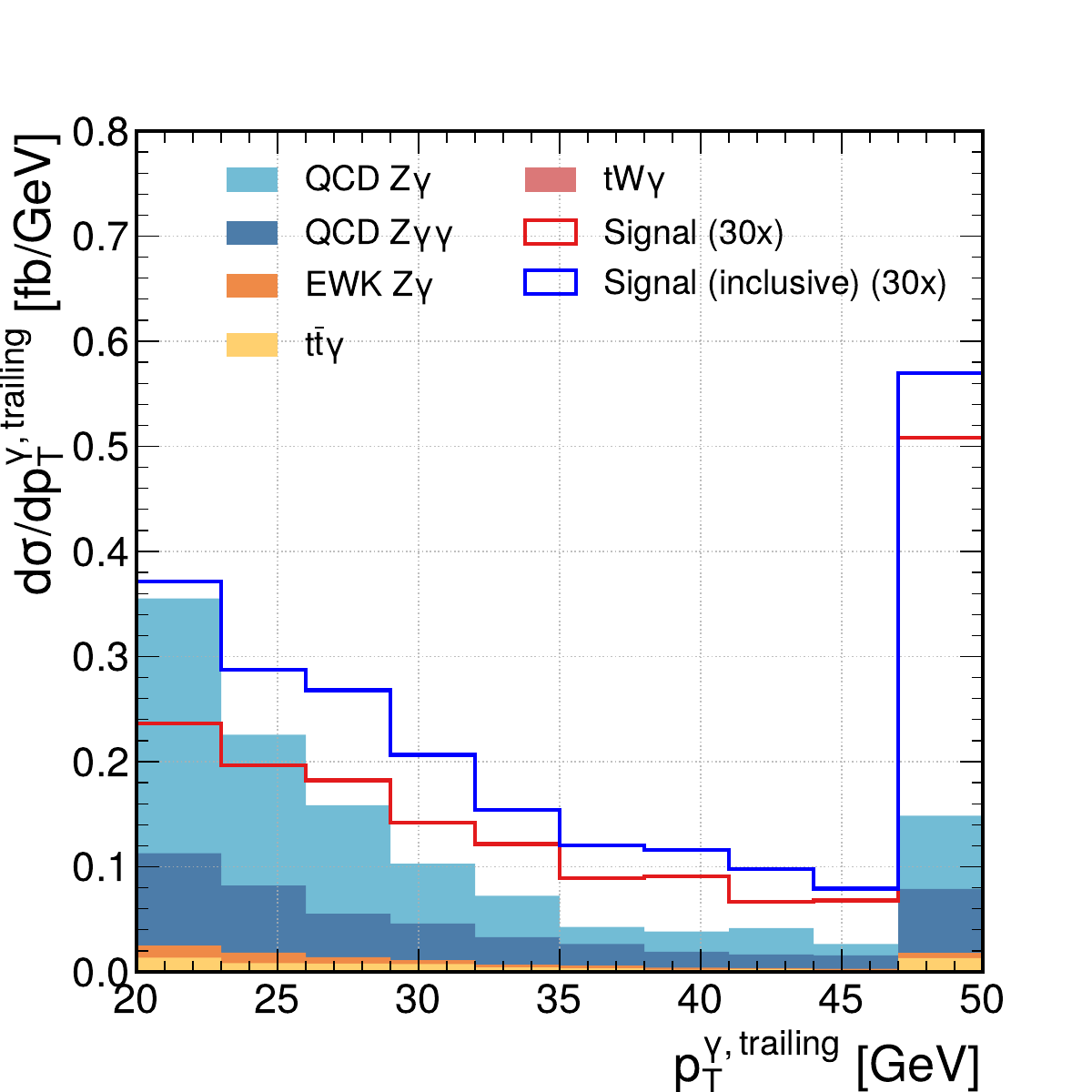}\label{fig:id1_2}}\\
    \subfloat[$M_{jj}$]
    {\includegraphics[width=.495\textwidth]{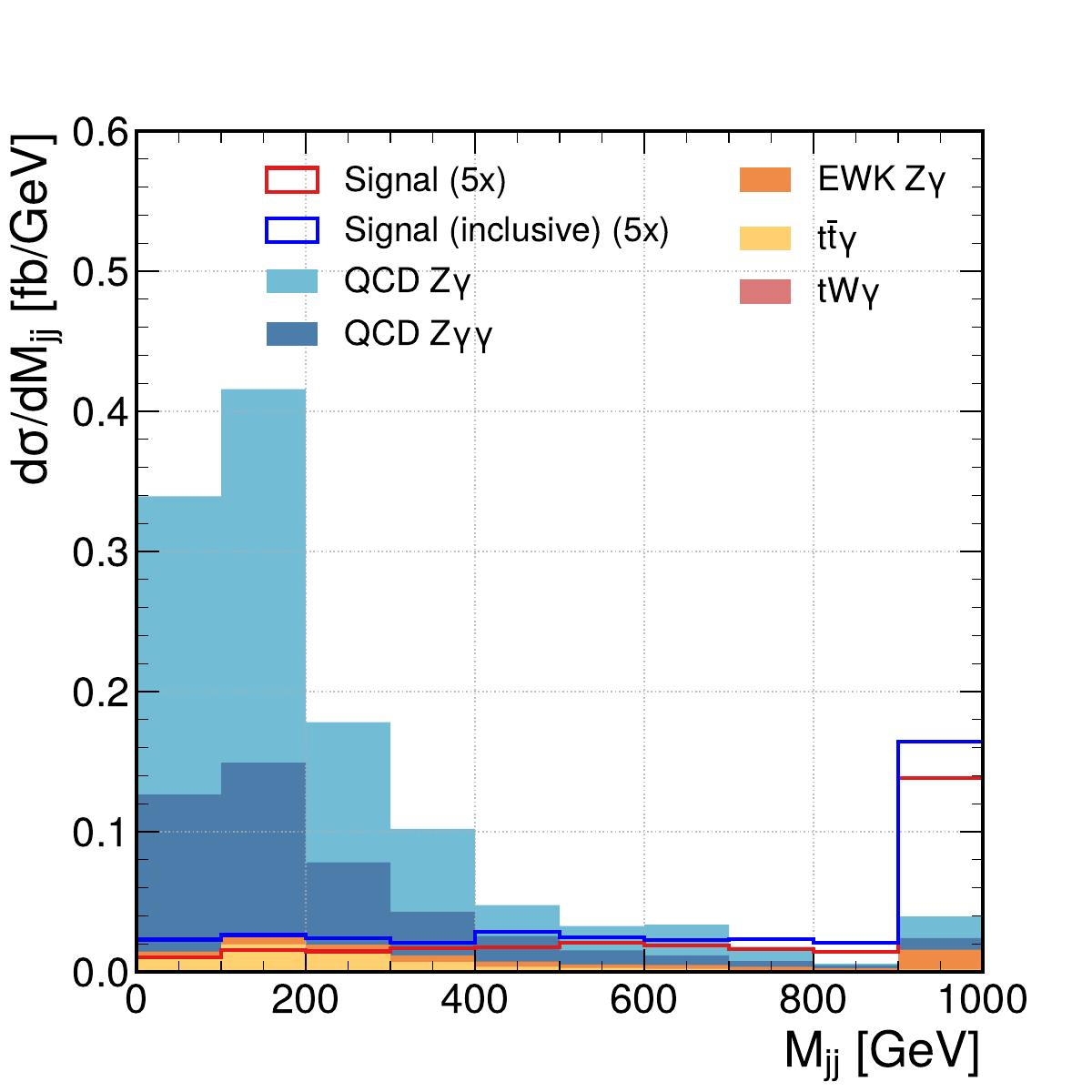}\label{fig:id1_3}}
    \subfloat[$\left|\Delta\eta_{jj}\right|$]
    {\includegraphics[width=.495\textwidth]{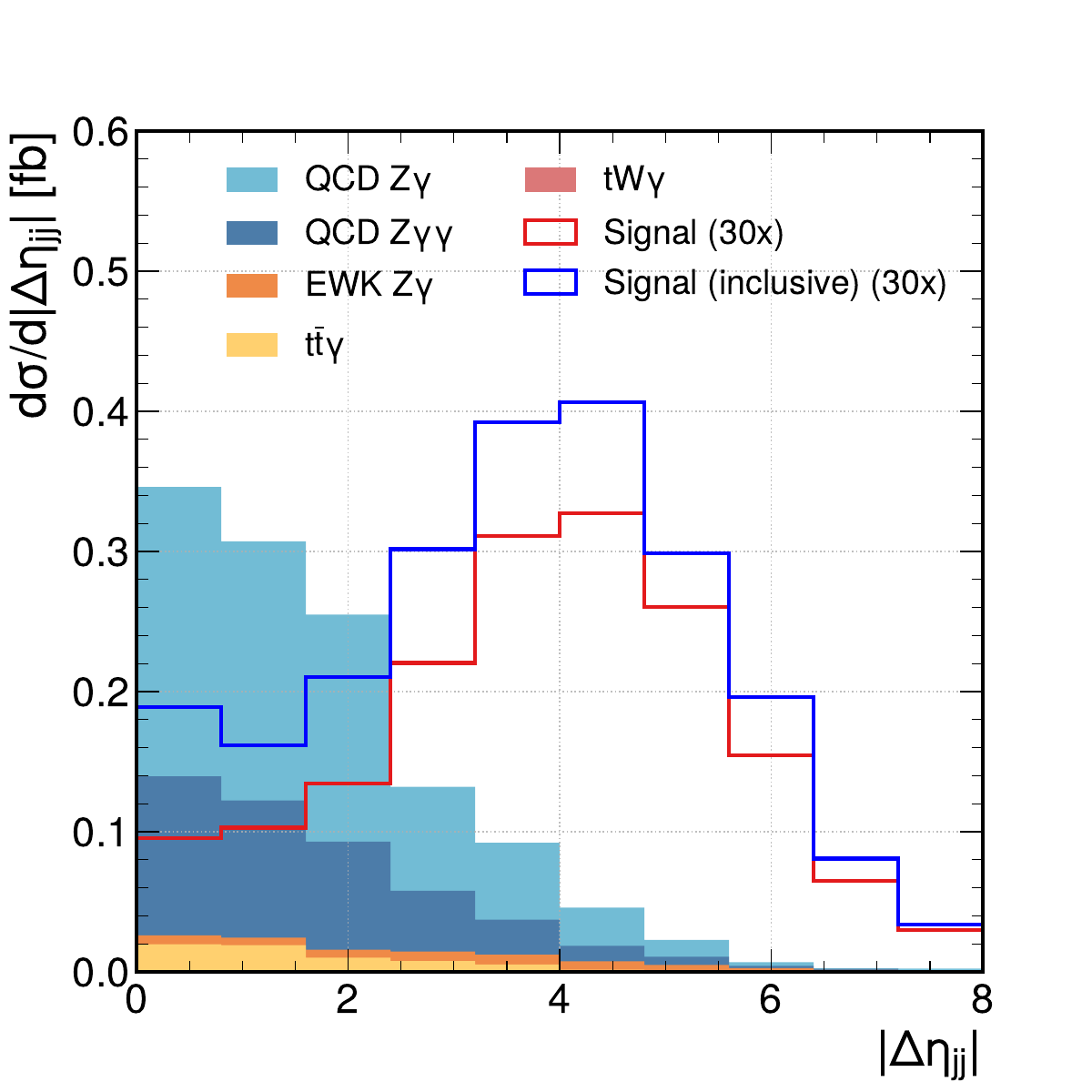}\label{fig:id1_4}}
    \caption{Distributions of signal and background after initial signal selection.}
    \label{fig:distribution_initcut}
\end{figure*}

\begin{figure*}[!h]
    \centering
    \subfloat[$M_{\ell^+\ell^-\gamma\gamma}$]
    {\includegraphics[width=.495\textwidth]{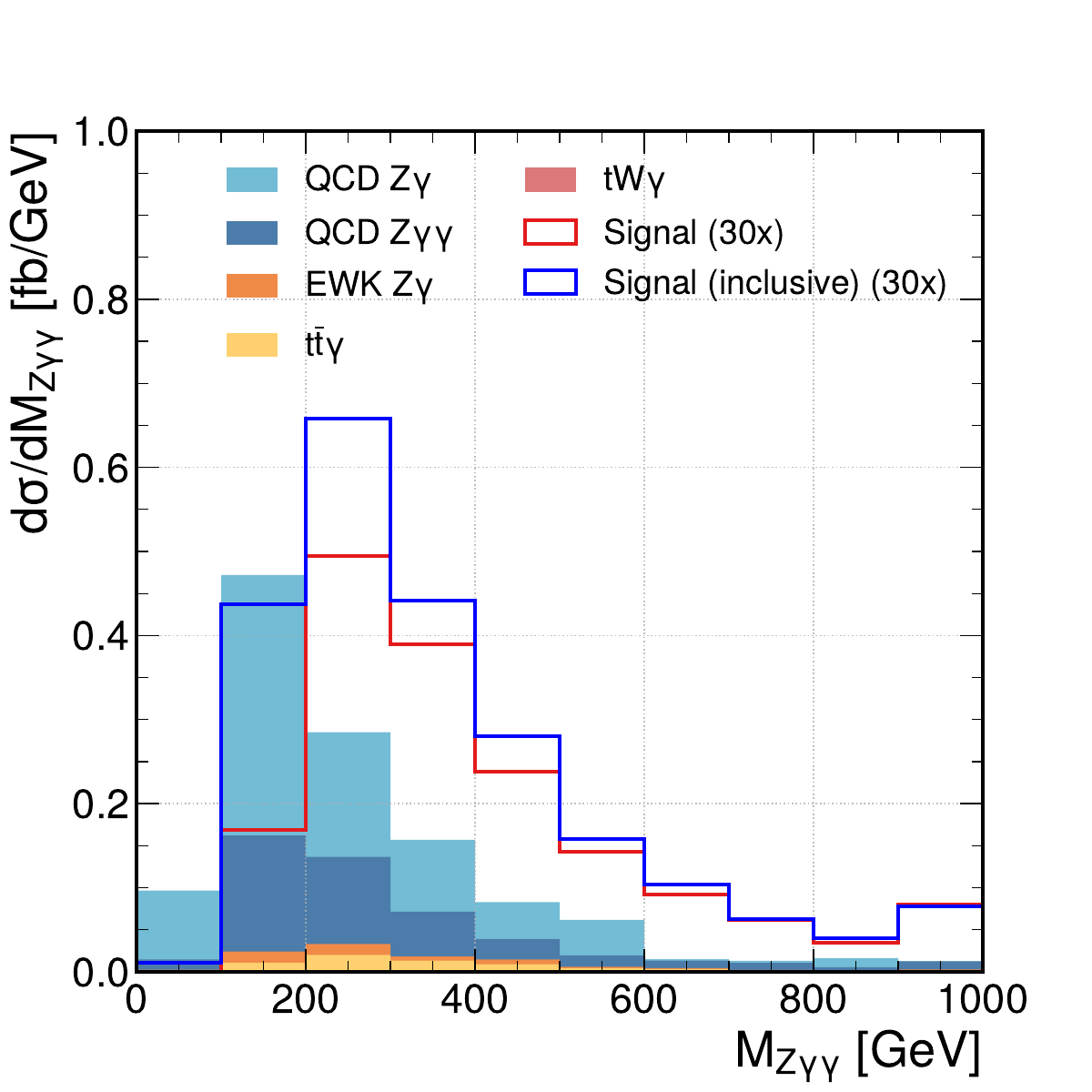}\label{fig:id1_5}}
    \subfloat[$p_{\text{T}}^{\gamma\gamma}$]
    {\includegraphics[width=.495\textwidth]{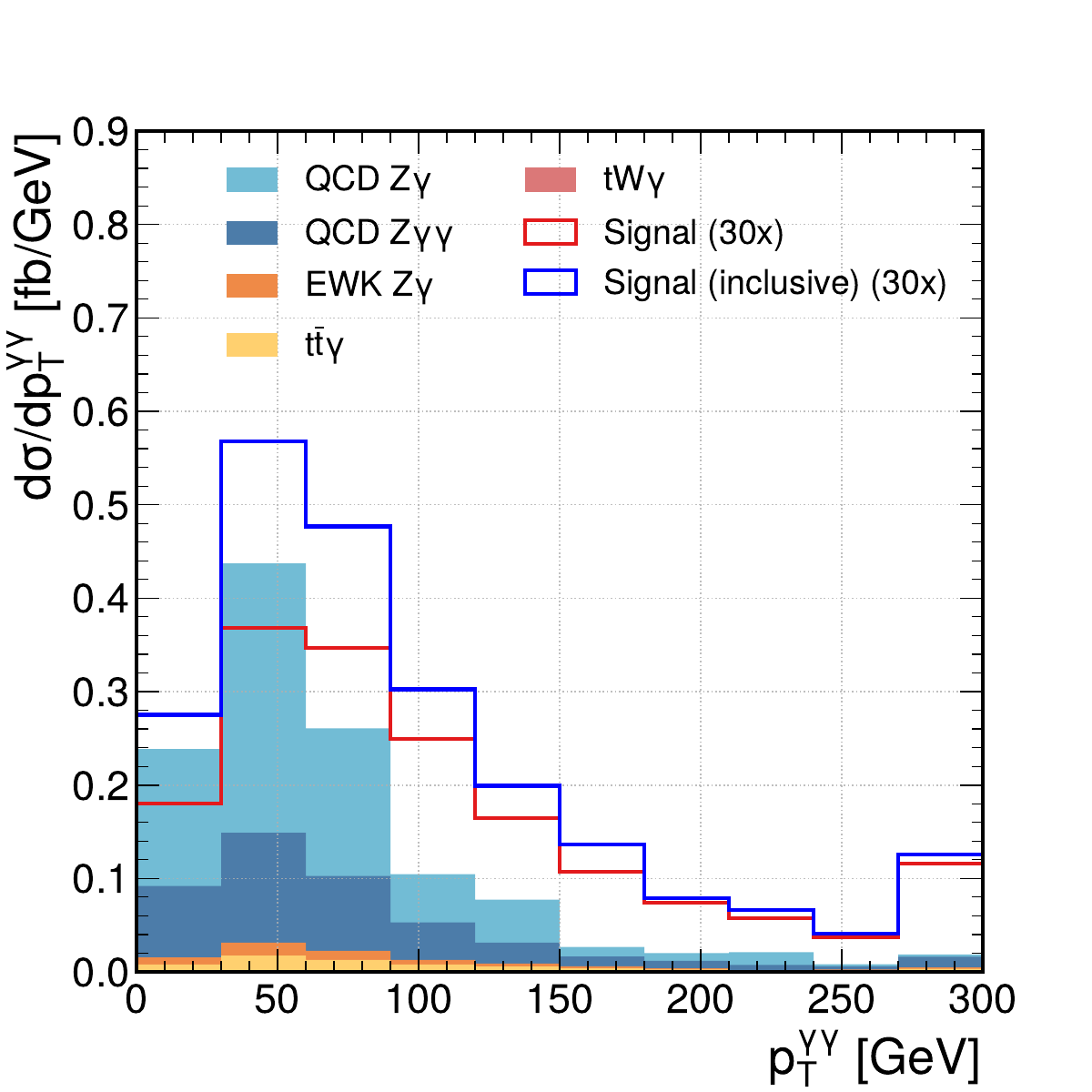}\label{fig:id1_6}}
    \caption{Distributions of signal and background after initial signal selection.}
    \label{fig:distribution_initcut2}
\end{figure*}

For events that pass the initial selection, the signal jets and photons are defined as the top two jets or photons with the highest $p_{\rm T}$. Meanwhile, the signal leptons pair is the one with the closest $M_{\ell^+\ell^-}$ to $M_Z$ (91.2~GeV). Then distributions of some variables are shown in Figure \ref{fig:distribution_initcut} and \ref{fig:distribution_initcut2}. All histograms, except for Figure \ref{fig:id1_1}, employ capping at the endpoints to retain outliers; Figure \ref{fig:id1_1}, in contrast, excludes events outside the plotted range. The per-event weight is calculating by $w_i = \sigma_i/N_i$, where $\sigma_i$ and $N_i$ are the cross-section and the number of simulated events of the corresponding process, respectively. Due to the low cross-section of the signal processes, they are scaled 5 or 30 times in the plots for better illustrating the shapes.

It shows that the most significant background contribution comes from QCD $Z\gamma$, while the QCD $Z\gamma\gamma$ background is also non-negligible. These backgrounds can be efficiently removed by applying a lower cut on $M_{jj}$. Furthermore, additional variables including $M_{\ell^+\ell^-}$, $\left|\Delta\eta_{jj}\right|$, and $M_{\ell^+\ell^-\gamma\gamma}$ may also be applied for background suppression.

\section{Events yield and significant study}
\label{sec:yieldandsignificance}

Applying $M_{\ell^+\ell^-\gamma\gamma}>100~\text{GeV}$ and $70~\text{GeV}<M_{\ell^+\ell^-}<120~\text{GeV}$ sequentially to events that pass the preliminary signal selection, the relative efficiency ($\epsilon_{\text{rel.}}$) and remaining cross-section ($\sigma$) for each step are shown in Table \ref{tab:efficiency_table}.

\begin{table}[htbp]
    \centering
    \renewcommand{\arraystretch}{1.3}
    \begin{tabular}{l c c  c c  c c}
    \toprule
	\midrule
    \textbf{\textbf{\quad Category}}  &
    \multicolumn{2}{c}{\quad \textbf{Initial Selection}\quad} &
    \multicolumn{2}{c}{\quad\textbf{$M_{\ell^+\ell^-\gamma\gamma}$ Cut}\quad} & 
    \multicolumn{2}{c}{\quad\textbf{$M_{\ell^+\ell^-}$ Cut}\quad} 
    \\ 
    \cline{2-7}
    &\quad $\sigma$ (fb)&$\varepsilon_{\text{rel.}}$ \quad& \quad$\sigma$ (fb)&$\varepsilon_{\text{rel.}}$&$\sigma$ (fb)&$\varepsilon_{\text{rel.}}$\quad\\ 
    \midrule
Signal&\quad5.66E-2\quad&\quad21.56\%&\quad5.66E-2&\quad100.00\%&\quad5.56E-2&\quad98.25\%\\
Signal (inclusive)&\quad7.56E-2\quad&\quad17.76\%&\quad7.52E-2&\quad99.51\%&\quad6.27E-2&\quad83.39\%\\
\midrule
QCD $Z\gamma\gamma$ &\quad3.76E-1\quad&\quad1.64\%&\quad3.64E-1&\quad96.82\%&\quad2.18E-1&\quad59.82\%\\
QCD $Z\gamma$ &\quad7.29E-1\quad&\quad0.01\%&\quad6.47E-1&\quad88.69\%&\quad3.04E-1&\quad47.06\%\\
EWK $Z\gamma$ &\quad4.54E-2\quad&\quad0.13\%&\quad4.31E-2&\quad95.04\%&\quad2.61E-2&\quad60.58\%\\
$t\bar{t}\gamma$ &\quad6.44E-2\quad&\quad0.19\%&\quad6.40E-2&\quad99.31\%&\quad1.76E-2&\quad27.48\%\\
$tW\gamma$ &\quad4.41E-5\quad&\quad0.19\%&\quad4.39E-5&\quad99.47\%&\quad7.43E-6&\quad16.93\%\\
Background in total &\quad1.22E+0\quad&\quad 0.01\%&\quad1.12E+0&\quad 92.01\%&\quad5.66E-1&\quad 50.62\%\\
	\midrule
    \bottomrule
    \end{tabular}
    \caption{Summary of the relative efficiency ($\epsilon_{\text{rel.}}$) and remaining cross-section ($\sigma$) after each selection or cut.}
    \label{tab:efficiency_table}
\end{table}

The background can be suppressed substantially by applying additional cuts on $M_{jj}$ and $\left|\Delta\eta_{jj}\right|$. Given the correlation between them, a two-dimensional parameter scan is adopted to determine the optimal selection criteria, with lower cut of $M_{jj}$ varying from 200~GeV to 800~GeV in steps of 100~GeV and $\left|\Delta\eta_{jj}\right|$ from 0 to 4.0 in steps of 0.5. The signal significance for each ($M^{\text{lower}}_{jj}$, $\left|\Delta\eta_{jj}\right|^{\text{lower}}$) combination, derived from the corresponding signal and background yields, is calculated by
\begin{equation}
    Z = \sqrt{2\left[(s+b)\ln\left(1+\frac{s}{b}\right)-s\right]},
\end{equation}
where $s$ and $b$ are the signal and the total background yield, respectively. Specially, the signal yield is from the Signal (inclusive) process. 

\begin{figure}
    \centering
    \includegraphics[width=0.95\linewidth]{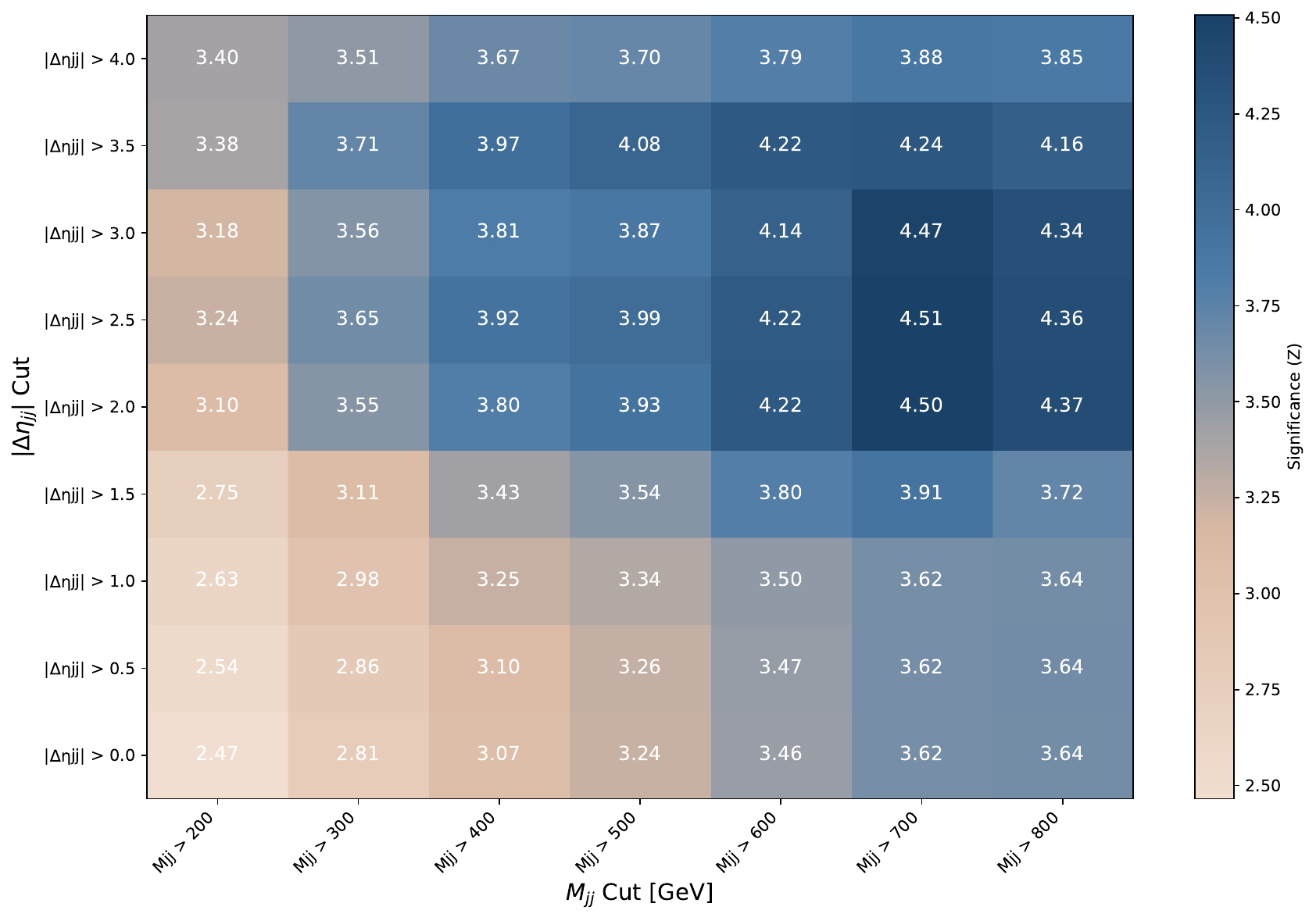}
    \caption{Significance for each ($M^{\text{lower}}_{jj}$, $\left|\Delta\eta_{jj}\right|^{\text{lower}}$) combination, derived from the corresponding signal and background yields ($L = 500~\text{fb}^{-1}$).}
    \label{fig:significance}
\end{figure}

The results with $L = 500~\text{fb}^{-1}$ is given in Figure \ref{fig:significance}. It shows a peak significance of 4.51~$\sigma$ at (700~GeV, 2.5). 
Due to the limited statistics of the simulation, the statistical uncertainties of $s$ and $b$ must be considered. These are calculated from the number or proportion of the remaining events, where the signal yield follows a binomial distribution, and the background processes can be further approximated by a Poisson distribution.
The statistical uncertainties of the regions among the peak significance are approximately around 0.2. Therefore, the variations between points are mostly within one standard deviation, when $M_{jj}^{\text{lower}}$ is set greater than 700 and $\left|\Delta\eta_{jj}\right|^{\text{lower}}$ is within the range of 2.0–3.0. While a definitive choice would require higher statistics, we select (700~GeV, 2.0) as the final cut in a conservative manner. Table \ref{tab:yield_table} shows the event yields of each signal and background processes after the full selections. The final significance is $4.50\pm 0.17~\sigma$ at $L = 500~\text{fb}^{-1}$. And for the High-Luminosity LHC~\cite{ZurbanoFernandez:2020cco} with $L = 3000~\text{fb}^{-1}$, the significance can exceed 10~$\sigma$.

\begin{table}[htbp]
    \centering
    \renewcommand{\arraystretch}{1.3}
    \begin{tabular}{ c  c c}
    \toprule
	\midrule
    \textbf{\textbf{\quad Category}} &\quad Event Yield \quad& \quad Uncertainty (stat.)\quad\\ 
    \midrule
\quad Signal\quad&16.28 &	0.27\\ 
Signal (inclusive)&17.55 &	0.46\\ 
\midrule
QCD $Z\gamma\gamma$&3.32 	&0.53 \\
QCD $Z\gamma$&1.52 	&0.54 \\
EWK $Z\gamma$&5.36 	&0.40 \\
$t\bar{t}\gamma$&0.15 	&0.10 \\
$tW\gamma$&0.00 	&0.00 \\
\quad Background in total\quad &10.36 & 0.86\\
	\midrule
    \bottomrule
    \end{tabular}
    \caption{Event yields of each process after the full cut flow ($L = 500~\text{fb}^{-1}$).}
    \label{tab:yield_table}
\end{table}

\section{Conclusion}
\label{sec:summary}
VBS is crucial for testing the SM, especially the EWK symmetry breaking mechanism. The ongoing release of the LHC Run 3 dataset presents opportunities to conduct searches for a series of $2\to 3$ VBS processes, with the prospect of breakthroughs.

This study presents an investigation of the EWK $\text{pp} \to Z\gamma\gamma + 2~\text{jets}, ~Z\to \ell^+\ell^- ~(\ell = e,\mu)$ process, with particular focus on the VBS-enriched region. Using MG5, PY8, and Delphes to perform MC simulation and detector fast simulation, we get the distributions of the signal and series of background processes, and thus develop optimized selection criteria to suppress the background.

The analysis demonstrates that requirements on $M_{jj} > 700~\mathrm{GeV}$ and $|\Delta\eta_{jj}| > 2.0$ are crucial for QCD-induced background suppression, yielding a signal significance of $4.50\pm 0.17~\sigma$ with an integrated luminosity of $500~\mathrm{fb}^{-1}$. The significance is projected to exceed $10~\sigma$ at the HL-LHC with $3000~\mathrm{fb}^{-1}$, highlighting the potential for observation and detailed study of this rare process in future LHC data. These findings establish a solid foundation for experimental searches and underscore the viability of $2\to3$ VBS processes.

\begin{acknowledgments}
This study is supported by the National Natural Science Foundation of China
under Grant No. 12325504. 
\end{acknowledgments}

\bibliography{main}

\end{document}